\documentclass[12pt, preprint]{aastex}
\usepackage{amsmath}
\usepackage{natbib}
\shortauthors{Strolger \& Riess}
\shorttitle{Supernovae in the Hubble Ultra Deep Field}

\newcommand\adeg{\arcdeg}
\newcommand\amin{\arcmin}
\newcommand\asec{\arcsec}

\begin{document}
\title{The Deepest Supernova Search is Realized in the Hubble Ultra Deep Field Survey
\footnote{Based on observations with the NASA/ESA {\it Hubble Space Telescope}, obtained at the Space Telescope Science Institute, which is operated by AURA, Inc., under NASA contract NAS 5-26555} 
}

\author{Louis-Gregory~Strolger and Adam~G.~Riess}

\affil{Space Telescope Science Institute, 3700 San Martin Drive, 
Baltimore, MD 21218\\{\tt strolger@stsci.edu, ariess@stsci.edu}}

\begin{abstract}
The Hubble Ultra Deep Field Survey has not only provided the deepest optical and near infrared views of universe, but has enabled a search for the most distant supernovae to $z \sim 2.2$. We have found four supernovae by searching spans of integrations of the Ultra Deep Field and the Ultra Deep Field Parallels taken with the {\it Hubble Space Telescope} paired with the Advanced Camera for Surveys and the Near Infrared Multi Object Spectrometer.  Interestingly, none of these supernovae were at $z > 1.4$, despite the substantially increased sensitivity per unit area to such objects over the Great Observatories Origins Deep Survey. We present the optical photometric data for the four supernovae. We also show that the low frequency of Type Ia supernovae observed at $z>1.4$ is statistically consistent with current estimates of the global star formation history combined with the non-trivial assembly time of SN~Ia progenitors.

\end{abstract}


\section{Introduction}
The Hubble Ultra Deep Field (UDF) survey (GO 9978, S. Beckwith, PI) is the deepest view of the universe in the optical and near infrared obtained to date.  Imaging was acquired with the Advanced Camera for Surveys (ACS) at a position located within the Great Observatories Origins Deep Survey~\citep[GOODS,][]{Giavalisco:2003ig} South Field, centered at R.~A.~(J2000)=03h32m39'00 and Decl.~(J2000)=-27\adeg47\amin29.1\asec. A total of 400 {\it HST} orbits were accumulated to achieve a $10\sigma$ detection threshold of approximately 29 mag (Vega) in the $F435W$, $F606W$, and $F775W$-bands,  and approximately 28 mag in the $F850LP$-band. To further enhance the vast array of deep multi-wavelength imaging for this region, deep near infrared images were also obtained in the UDF target field (IRUDF; GO 9803, R. Thompson, PI) by acquiring individual pointings with the Near Infrared Camera and Multi Object Spectrometer (NICMOS) camera 3, each 8 orbits in depth in both the $F110W$ and $F160W$-bands, to cover a 3 x 3 mosaic near the center of the UDF. This observation was divided into two epochs separated by 90 days to enable searches for SNe.  Another extremely deep optical field, an ACS ``parallel field'', observed during the imaging of the IRUDF, overlapped with GOODS South observations. The UDF, the UDF ACS Parallel (UDFP), and the IRUDF surveys are unique in that they are 
the first surveys sensitive to Type Ia supernovae (SNe~Ia) to $z \le 2.2$.

With an elasped time of less than 4 Gyr between the first generation of stars ($z \sim 100$) and SNe Ia at $z > 1.4$,  the observed number of these events at $1.4 < z < 2.2$ could provide an important probe of the assembly time required by SNe Ia progenitors~\citep[][hereafter S04]{Strolger2004a}. The rate in which SN Ia events occur is governed by the rate in which their progenitor stars form, and the time  required for the SN Ia progenitor to develop into a SN Ia event. In the framework of cosmic time, these components are the star formation rate history, and the population assembly time, or the delay time function~\citep[][S04]{1999AA...350..349D}. Together, the star formation rate history and the delay time function describe a model for the SN Ia rate history [$\mathcal R_{Ia}(z)$] , which can be compared to observations of the SN Ia rate in different redshift regimes. 

In principle, the observed discovery rate of SNe at $z>1.4$ could help to distinguish between viable models of $\mathcal R_{Ia}(z)$. In S04, it was shown that observations of the SN Ia rate over a large redshift range (encompassing the most of the last $\sim10$ billion years of the universe) are consistent with the combination of the star formation rate history from rest-frame $U$-band galaxy studies [hereafter SFR$_U$($z$), see~\citet{Giavalisco:2003bi}], and a Gaussian distribution of delay times, with a mean $\sim 4$ Gyr and a dispersion of $\sim 1$ Gyr.

However, recently it has been shown that the rate SNe~Ia per galaxy  at $z < 0.1$ does appear to increase towards later galaxy types, e.~g.~the SN Ia rate in irregular (dwarf) galaxies is approximately 10 times larger than in elliptical galaxies of the same total mass~\citep{mannucci}. This would seem to imply that the SN Ia rate more closely traces the star formation rate history rather than being largely delayed from it. This is difficult to resolve in light of the GOODS supernova data. However, one could hypothesize that the rest-frame $U$-band galaxy observations (including dust corrections), in reality, provide a rather poor tracer of actual star formation history, or only a lower limit~\citep{2001ApJ...556..562C}. Or, perhaps the completeness thresholds for the GOODS SN survey, the sensitivity as a function of change in flux, were significantly overestimated. 

A simple test of the S04 model would be to push the limits of survey sensitivity well beyond those achieved in the GOODS survey, and continue to look for high redshift SNe Ia at $z > 1.4$. One can then compare observed yields to those expected in this redshift regime from the S04 and other viable models of the SN~Ia rate history. If $\mathcal R_{Ia}(z)$ is indeed delayed by $\sim 4$ Gyr from the SFR$_U$($z$), then it is expected that the rate at $z> 1.4$ will be very low, and thus essentially no ($\ll 1$) SNe~Ia in this range could be found in this deep survey. Even if the $\mathcal R_{Ia}(z)$ is closely tied to the SFR$_U$($z$), i.~e.~with mean delays of less than approximately one billion years from star formation, then there could be on the order of only one SN~Ia discovery at $z> 1.4$. However, a very meaningful test this survey could provide is if the $\mathcal R_{Ia}(z)$ greatly increases with redshift, as it would if it were tied to the SFR$_{IR}$($z$), where one could expect at least few SNe~Ia  at $z > 1.4$ in this deep field survey.

We have searched spans of images of the UDF, UDFP, and IRUDF observations for high redshift SNe and have found a total of four events, none of which were at $z>1.4$. In \S 2 we describe the UDF+UDFP and IRUDF searches, show discovery and subsequent photometric data for the events found in the optical data, and present the evidence which show the relatively low redshift of these events. In \S 3 we compare the observed low observed yield at $z> 1.4$ to that which would be predicted from a few SN rate models.

\section{The Searches, Discovery Data, and ACS Photometry}\label{sec:SNe}

The UDF and UDFP images were searched by differencing stacks of images in the $F850LP$-band, assembled as they were collected over the duration of the deep survey. A total of nine multi-orbit epoch image stacks were created for the UDF field, and two were made for the UDFP. The first stack for each field was differenced with an overlapping area in either the ACS GOODS mosaic in the case of the UDF, or with ACS images obtained for GO~9352 (A. Riess, PI) in the case of the UDFP. The subsequent stacks in both surveys were differenced with their preceding stacks. Similarly, the IRUDF images were searched by differencing two epochs of mosaic images, each 4 orbits in depth, in both the $F110W$ and $F160W$ passbands. The two mosaic image stacks of the IRUDF were separated by approximately 77 days. In each the UDF, UDFP, and IRUDF searches, image stacks were produced using {\it multidrizzle}~\citep{koekemoe2003} by using the median of several exposures. This successfully rejects cosmic rays without rejecting transients which last substantially longer than the length of one exposure ($\sim1215$ seconds). The UDF and UDFP were drizzled to $0.''05$ per pixel in these preliminary image stacks, and later re-drizzled to $0.''03$ per pixel for the final release of the full-depth images. The IRUDF was drizzled to $0.''09$ per pixel, chosen to be a multiple of the final UDF pixel scale.

Subtracted images were mined for candidate SNe by automated computer searches and close visual inspection of the images. The criteria for identifying potential SNe, and for rejecting possible confusion sources, were very similar to those used in S04. Each image stack of the UDF and UDFP would be sensitive to motions only larger than the point spread function ($\approx0''.1$ FWHM in $F850LP$) over the accumulation time of the image stack (roughly one week). The proper motion of any candidate would have to be less than $6\times10^{-4}$ arcseconds hr$^{-1}$ ($1.4\times 10^{-3}$ degrees yr$^{-1}$) to be misidentified. This would exclude any foreground solar system object, as it would need to be farther than $247,000$ AU if orbiting the Sun at $\sim30$ km s$^{-1}$, which is well beyond the current estimates of the extent of the solar system. Slow-moving Galactic stars could also be generally excluded on similar hypothetical grounds. Variable stars are also easily excluded by their nature of being stars in all epochs of observation, and by generally appearing unassociated with galaxies in the field. Variable stars that, on occasion,  do appear aligned with field galaxies are typically several orders of magnitude brighter than one would expect for SNe in these galaxies.

What remained in the subtracted images are true extragalactic transients (e.~g.~supernovae and active galactic nuclei) and image artifacts such as misregistrations and variations in the point spread function. 

As the field is fairly dense with galaxies, and well resolved in the ACS images, there were no significant local misregistrations in the UDF and UDFP, except for stellar diffraction spikes which rolled with the change in orientation between epochs of the surveys. The point spread function (PSF) in each median combined image stack was largely stable and unaffected by short-term, exposure-to-exposure changes in the PSF (e.~g.~telescope breathing or focus drift). Thus, we are confident that we have identified only extragalactic transients from the UDF and UDFP subtraction images.

Variable active galactic nuclei (AGN) were defined as candidates that were within one pixel of their host nuclei. This is a rather conservative definition, as in principle, we would be capable of identifying transients with centroids a fraction of a pixel offset from the centroid of light in the nucleus of a galaxy. However, there were no detected candidates (down to the 5$\sigma$ threshold) that were $\le0''.05$ from the centroid of the host galaxy the UDF and UDFP searches. Other investigations which probe for much smaller optical variations (1 to 3$\sigma$) in galaxy nuclei do show evidence for low-level AGN in the UDF target field (Windhorst et al., in preparation), none of which would be undeniably identified with our differencing method.

The NICMOS images, however, suffered from intrapixel sensitivity effects. Camera 3 is poorly sampled at $\sim 0''.2$ per pixel, and can produce as much as a 30\% flux variation depending if the peak of a PSF is centered on a given pixel, or is offset towards the edges of the pixel~\citep{storrs1999,1999PASP..111.1434L,2000adass...9..521H}. This flux variation in the pixel response function (PRF) is not removed in the flat-field correction, and imposes an inherent limitation in these images. With many dithered exposures of the same field, it is possible to map the PRF~\citep{storrs1999,1999PASP..111.1434L}, and determine a correction to PSF photometry. Unfortunately, there are no easy means to deconvolve the PRF from the PSF. Therefore, differencing the two image stacks of the IRUDF necessarily resulted in a ``checkerboard'' pattern of under and over subtractions (measured at about $\pm 15 - 20\%$). Faint and diffuse objects were generally cleanly subtracted.

In Tables~\ref{tab:tab3} and~\ref{tab:tab4} we list the epoch stacks created for the UDF, UDFP, and IRUDF searches, with the mean date of the image stack, the total number of orbits, and the total exposure time of each stack. To asses the sensitivity and completeness of the UDF, UDFP, and IRUDF searches, we performed monte carlo tests with planted PSFs meant to represent false SNe.

 In the case of the IRUDF, a PSF was generated from a few stars in the images, and scaled to one count per second using zero points and aperture corrections provided by the NICMOS group at Space Telescope Science Institute. The measured aperture photometry was then corrected for intrapixel sensitivity variations following a prescription detailed in~\citet{storrs1999}.

\begin{figure}
        \epsscale{0.8}
        \plotone{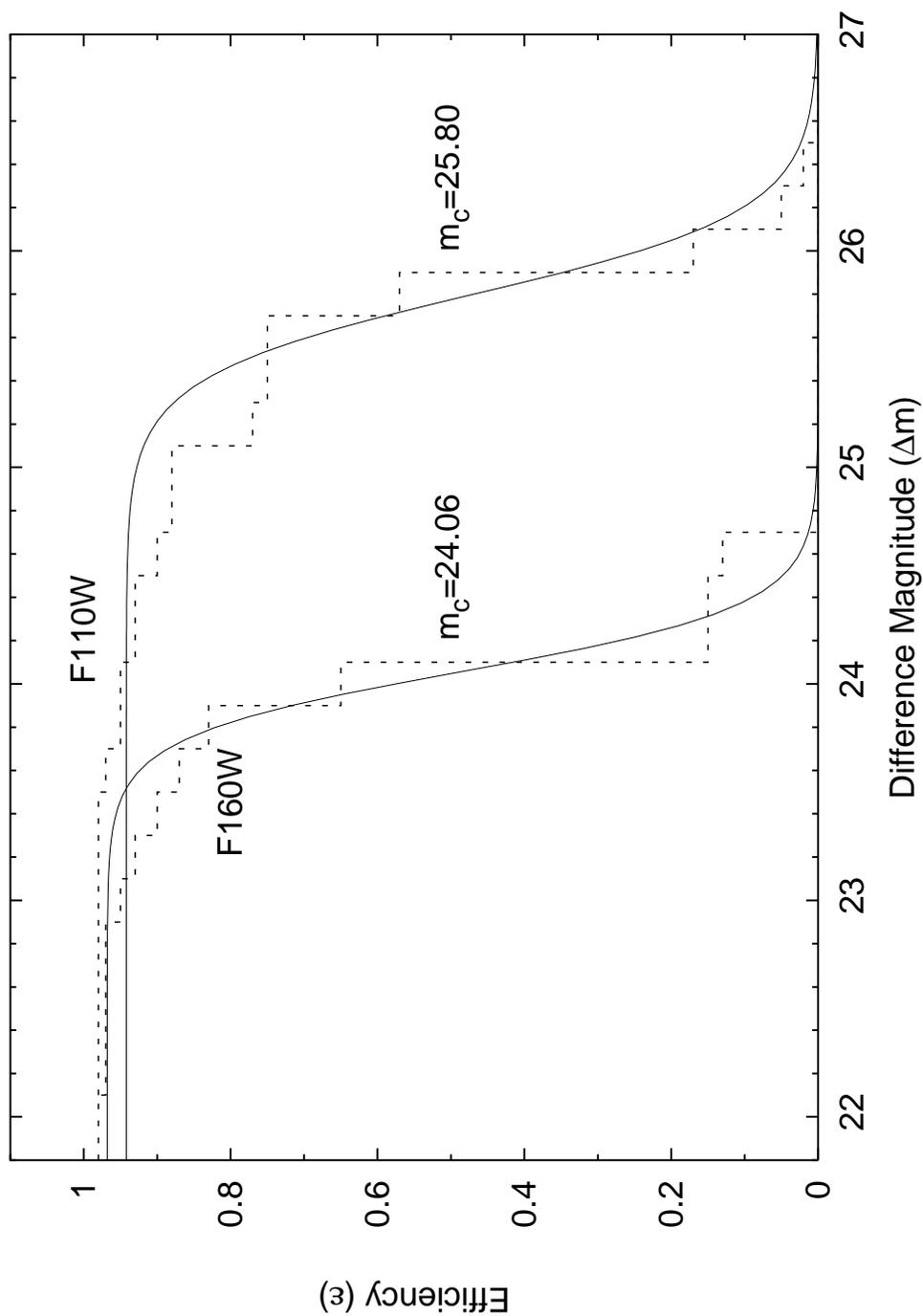}
        \caption{Efficiency of the IRUDF survey in recovering false SN placed in the centers of galaxies in the field. The fraction of recovered fake SNe per magnitude bin is shown as a histogram in the $F110W$ and $F160W$ passbands, and is approximated by the function $\varepsilon(\Delta m)\propto (1+e^{\Delta m})^{-1}$.\label{fig:fig1}}
\end{figure}

We placed one false SN of a given magnitude on the centers of a randomly selected set of 50 detected objects (to $>5\sigma$), subtracted the images, and then attempted to recover the false SNe by visual inspection (without prior knowledge of the locations of each planted false SN). This method was iterated to successively fainter magnitudes (in steps of 0.2 mag) until none of the planted SNe were recovered. The resulting histograms of percent recovered per magnitude bin is shown in Figure~\ref{fig:fig1}. 

The sensitivity in each epoch of the IRUDF was determined from the difference magnitude, $\Delta m$, determined from the flux difference of objects in the residual frame, which is given by:

\begin{equation}
        \begin{split}
     m_{1}&=ZP - 2.5\times \log(F_{1}) \Rightarrow F_{1}=10^{-\frac{2}{5}(m_{1} - ZP)}\\     
     m_{2}&=ZP - 2.5\times \log(F_{2}) \Rightarrow  F_{2}=10^{-\frac{2}{5}(m_{2}-ZP)}\\ 
     \Delta m&=ZP - 2.5\times \log(F_{2}-F_{1}),
     \end{split}
 \end{equation}

\noindent where $F_1$ and $F_2$ are the flux of the false SN in first and second image stacks, $m_1$ and $m_2$ are the corresponding magnitudes, and $ZP$ is the photometric zero point (total flux of 1 count per second). We use an analytical function to describe the efficiency in detecting objects of a given difference magnitude:

\begin{equation}
        \varepsilon (\Delta m) = \frac{T}{1 + e^{(\Delta m - \Delta m_{c})/S}},
        \label{eqn:efficiency}
\end{equation}

\noindent where $T$ is the maximum efficiency, $\Delta m_{c}$ represents a cutoff magnitude where
$\varepsilon(\Delta m)$ drops below 50\% of $T$, and $S$ controls the shape of
the roll-off. We find that $T=0.98$, $m_c=25.60$, and $S=0.20$ well describes the efficiency histogram for the $F110W$ passband, and $T=0.97$, $m_c=24.20$, and $S=0.21$ parameterizes the efficiency in the $F160W$ passband (see Figure~\ref{fig:fig1}).

\begin{figure}
        \epsscale{1.0}
        \plotone{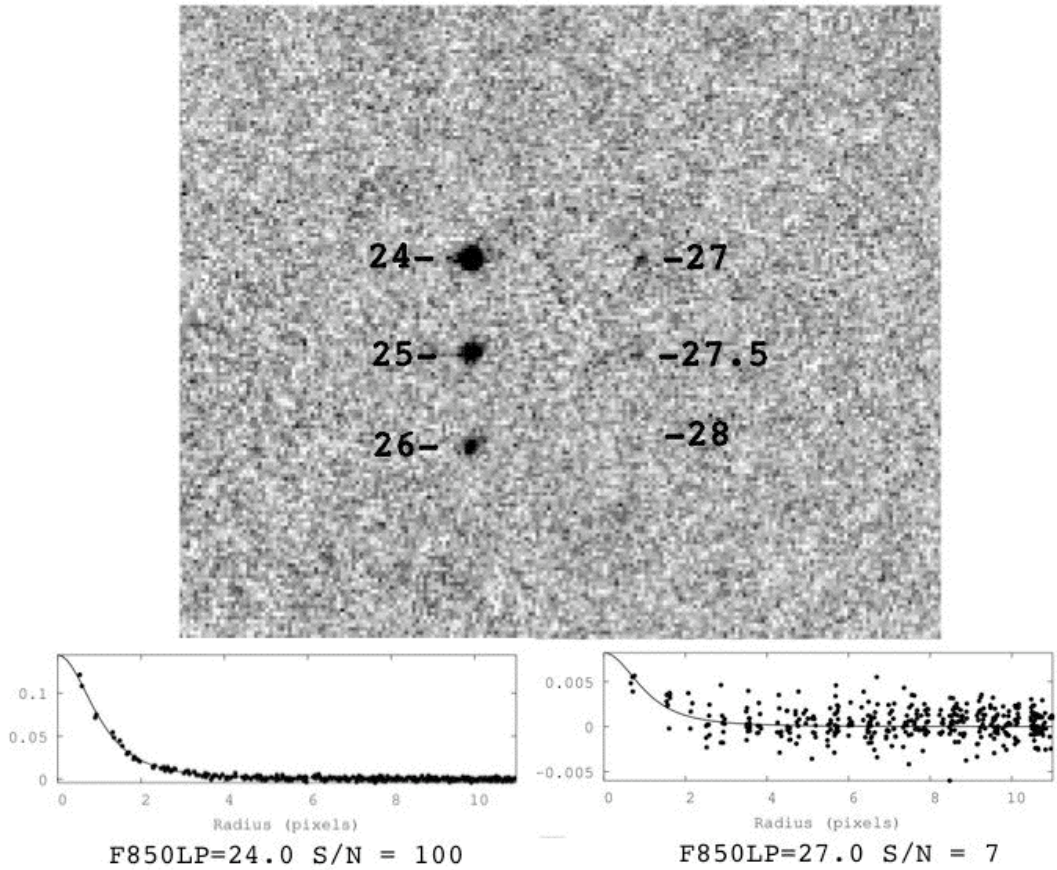}
        \caption{Example of the average survey sensitivity in the $F850LP$-band. A region of the 1224-1230 image stack is shown after differencing it with the previous image stack. Fake SNe added at various magnitudes. Below the image are radial profiles (solid points, lines are Moffat fits to the profiles) of the 24 mag and 27 mag fake SNe, with approximate S/N of 100 and 7 respectively. SNe above $\approx27.1$ mag (which corresponds to the $5\sigma$ limit) were virtually undetectable in this survey.\label{fig:fig2}}
\end{figure}

In the $F850LP$ passband of the UDF and UDFP surveys, false star tests show nearly identical efficiency function parameters as those used in S04, with $T=1$ and $S=0.38$, however with adjusted $50\%$ efficiency cutoff magnitudes ($m_c$) corresponding to the $5\sigma$ sensitivity limits for the difference of a given pair of stacks (also shown in Tables~\ref{tab:tab3} and~\ref{tab:tab4}). These limits are in good agreement with those expected from the exposure times of the search and template images using the ACS Exposure Time Calculators. The typical brightness threshold of detection for the $F850LP$ template-search pairs was $27$ mag. An illustration of the sensitivity threshold can be seen in Figure~\ref{fig:fig2} where we have added six fake SNe (PSFs) at $F850LP=$ 24, 25, 26, 27, 27.5, and 28 mag in a region of the 1224-1230 image stack. By differencing it with the 1212-1218 image stack, we can clearly detect the sources to a flux level of 27 mag, but fainter than 27 mag was difficult to identify without prior knowledge of where the fake SNe were.  This is in good agreement with the estimated detection threshold (from the exposure times) of 27.1 mag for this template-search pair. The detection thresholds are bright enough to detect SNe~Ia at peak at $z\le2.2$ in each passband.

\begin{figure}
        \epsscale{1.0}
        \plotone{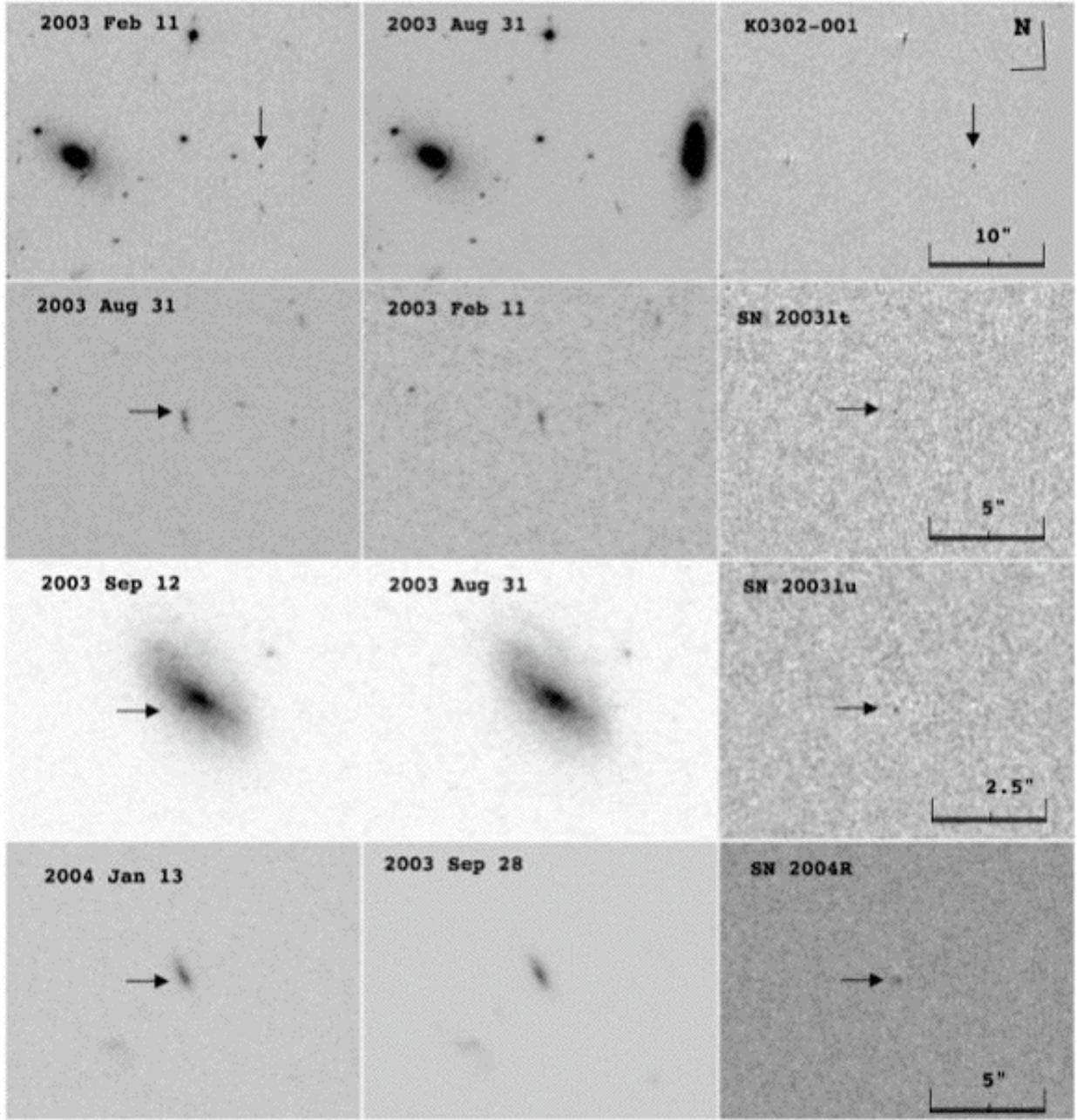}
        \caption{Discovery images for SNe~K0302-001, 2003lt, 2003lu, and 2004R (first column), shown with template images without the supernovae (middle column), and subtracted residual frame (last column). North is up and East to the left in all images.\label{fig:fig3}}
\end{figure}

Despite the intrapixel sensitivity limitations of the IRDUF survey, interestingly, no candidate SNe were discovered in the deep IRUDF imaging. However, four SNe were discovered in the UDF and UDFP images. For each, vega based aperture magnitudes were measured from difference images in the $F606W$, $F775W$, and $F850LP$ bandpasses, using aperture corrections and photometric error estimations described in S04. In all cases, we used photometric redshifts (phot-$z$) determined from the multi-wavelength GOODS data to estimate the redshifts of the host galaxies~\citep{Mobasher:2003bm}. Spectroscopic confirmation has not been obtained for these SNe. We also generally lack the photometric data and age constraints necessary to assuredly identify the SN types using the identification confidence scheme detailed in S04 and color selection methods described in~\citet{Riess:2003gz}. However, we can still use these techniques to reject combinations of SN type and redshift space, specifically we can reject the faint and red signature of SNe~Ia at very high redshifts. The discovery images for all four SNe are shown in Figure~\ref{fig:fig3}, and the positional and photometric data are listed in Tables~\ref{tab:tab1} and~\ref{tab:tab2}.

Three SNe were identified in the area of the UDFP field. K0302-001\footnote{This designation reflects the new {\it International Astronomical Union} (IAU) standard for possible faint supernovae. See~\url{\tt http://cfa-www.harvard.edu/iau/CBAT\_PSN.html}} was discovered in the GOODS follow-up images from GO 9352, and not detected seven months later (to within 5$\sigma$) in the UDFP images. The host of K0302-001 was very faint, with $F606W\approx 28$ mag within a 0.2\arcsec~radius, and very blue, virtually undetectable at $F850LP\ge27.5$ in the same aperture. The galaxy could not be sufficiently detected in any of the deep multi-wavelength ground-based data sample assembled for the GOODS (spanning $U$ though $K_s$-bands), and was only identified in two ACS bands: $F606W$ and $F775$. The lack of photometric measurements of the host galaxy made it difficult to constrain its photometric redshift using the Bayesian Photometric Redshift (BPZ) method~\citep{2000ApJ...536..571B}.   The phot-$z$ estimate derived of the host of K0302-001 lacks a significant peak and has a broad 95\% confidence interval of $1.03<z<2.22$.  However, the photometry of the SN is more illuminating.  The magnitude and colors (specifically the red $F606W-F850LP$ color of 3.33 mag; see \citeauthor{Riess:2003gz}~\citeyear{Riess:2003gz}) match those of the five SNe~Ia measured by~\citet{Riess:2004b} at or near maximum at $z\approx1.3$.  Results from~\citet{Riess:2003gz,Riess:2004b} demonstrate that $\sim$ 95\% of SNe with these photometric characteristics are correctly identified as SNe Ia (with the other 5\% being SNe Ic or Ib). A firm conclusion we can make is that this SN is too bright and too blue to be a SN Ia (of any previously seen luminosity and color) at $z > 1.4$.  

SN~2003lt was discovered in the first UDFP stack (mean date 2003 Aug.~31),  and undetected in the GOODS follow-up comparison images. It was also well detected in the second UDFP stack of mean date 2003 Sep.~12. The host galaxy was well detected in several passbands, and thus the phot-$z$ for the host was well constrained at $z=1.0$ ($0.74<z<1.26$ 95\% confidence interval). We find that the photometry and colors of this SN were consistent with those for a SN~Ia discovered $\sim 80$ days from maximum light at $z=1.0$, and inconsistent with tested SN Ia scenarios (varying age, light-curve shape, and extinction) at $z>1.4$.

SN~2003lu was found in the second UDFP stack, and not detected in the first UDFP stack. The single $F850LP$-band measurement alone does not allow for a restriction in SN type and redshift space. However, the phot-$z$ for the bright and well-detected host was $z=0.11$ (0.0 - 0.25 95\% confidence interval), and therefore we can reject the possibility that this was a SN~Ia at $z>1.4$.

The only SN detected in several epochs of the UDF target field observations was SN~2004R, discovered in the last 6-orbit stack obtained 2004 Jan. 13. It was not detected, to within 5$\sigma$, in the 12-orbit stack from 2003 Dec. 24 - 30. A review of the data obtained 2004 Jan. 1 - 11 revealed that the SN was rapidly rising in the $F850LP$ and $F775W$-bands, but declining in the $F606W$-band. The phot-$z$ for the host galaxy was again well constrained at $z=0.8$ ($0.56< z< 1.04$ 95\% confidence interval). The light curves and colors were generally consistent with our Type II-Plateau model, SN~1999em, caught prior to rest-frame $B-$band maximum light, with a host extinction of $A_{F850LP}\approx1.5$ mag (assuming a Galactic extinction law) at the photometric redshift. The colors and magnitudes of the SN are inconsistent with SN Ia models at $z>1.4$.

To summarize, our searching of the UDF, UDFP, and IRUDF resulted in four SNe of largely unknown types, but we can reject the possibility than any were SNe Ia at $z>1.4$.

\section{Comparisons to Predicted Rate Models}\label{sec:results}

The sample of SNe~Ia from the GOODS has shown for the first time a distinct rise in the SN~Ia rate from $0.5<z<1.0$~\citep{Dahlen2003}. These measurements are statistically inconsistent with a constant (or nearly constant) $\mathcal R_{Ia}(z)$, as inferred from measures of the SN rate at $z<0.5$ ~\citep{Blanc:2004ws}. 

An apparent discrepancy exists in the SN~Ia rate measurements at $z \sim 0.5$.  At this redshift, the SN Ia rate measured from the GOODS data are nearly twice as high as measurements made in the ground-based high-$z$ supernova search programs~\citep{2002ApJ...577..120P,2003ApJ...594....1T}. The natural explanation would appear to be that completeness was not a primary goal for these ground-based programs, and was generally sacrificed for sake of favorable SNe~Ia with low background contamination for optimal spectroscopic confirmation and precise distance measurements. Indeed, a recent, careful  re-examination of the data from the Fall 1999 campaign of the High-$z$ Supernova Search project~\citep{2003ApJ...594....1T}, and the Fall 2001 IfA survey~\citep{Barris:2003dq} shows preliminary evidence of many additional SNe resulting an increase by a factor of two in the SN~Ia candidates, and thus a likely increase in the rates determined from these data to values consistent with the GOODS measurements (Brian J. Barris and John L. Tonry, in private communications). 

Perhaps the most intriguing result on the supernovae rates from the GOODS sample has been the dearth of SNe~Ia discovered at $z >1$. \citet{Dahlen2003} have found that there is a steep decline in the rate at $1.2< z <1.6$ in comparison to the rate at $z=1.0$, a result seemingly at odds with the lack of such a decline in the global star formation rate in this redshift range. However, S04 and~\citet{Dahlen2003} have shown that $\mathcal R_{Ia}(z)$ is a reflection of the SFR$_U$($z$) which has been delayed by the assembly time function (or delay time function), i.~e.~the time for SN~Ia progenitors to go from formation to explosion.   S04 find that the difference in the evolution of the SFR$_U$($z$) and the $\mathcal{R}_{Ia}(z)$ can be well modeled as a Gaussian delay time function with a mean delay of $\sim 4$ Gyr and a dispersion that is a small fraction of the delay. Subsequently~\citet{Dahlen2003} find that based on the model delay, the fraction of stellar mass in the range of 3 to 8 solar masses which will explode as SNe Ia is about $5\%$. 

Past searches for SNe Ia in high-redshift clusters with HST by~\citet{2002MNRAS.332...37G} show rates 
that are similar to field SN Ia rates, indicating some similarity in their rate evolution. Unfortunately, these cluster rate measurement at $z\gg0.1$ are not precise enough, nor sufficiently well sampled in redshift space, to definitively compare their SN~Ia rate history to those measured in normal (field) environments.

An anecdotal anomoly in the consideration of the decline in $\mathcal R_{Ia}(z)$ at $z>1$ is the serendipitous discovery of SN~1997ff, found in only one re-observation of a pointing with {\it HST} and WFPC2~\citep{Gilliland1999,2001ApJ...560...49R}. The combined depth of the template-search pair for the SN~1997ff discovery image was only marginally deeper than for a single tile of template-search epoch of the GOODS. The SN itself, although likely magnified by 0.3 mag~\citep{2002ApJ...577L...1B}, was discovered well above the detection threshold for the tiny survey.  The 5 month durration of the UDF+UDFP search and the factor of 2 increase in area of ACS over WFPC2 suggest that the number of such SNe Ia at $z>1$ found in the UDF should exceed those found by Gilliland et al. (1999) by a factor of $\sim$5.  Yet, when dealing with such small, pencil-beam surveys, the survey yields can be subject more to chance than expectations. The low number of expected SNe serve more as a probability for finding SNe, governed by Poisson statistics. 

In the previous section we concluded that our searches of the UDF, UDFP, and IRUDF had failed to yield
any SNe Ia at $z>1.4$.   To test the significance of this null result, we have simulated the yield expected in the multiple campaigns (image stacks) of the survey based on the best-fit $\mathcal R_{Ia}(z)$ determined in S04. As in S04, the number of expected SNe per redshift interval was determined by:

\begin{equation}
        N_{Ia}(z)=\mathcal R_{Ia}(z) t_c(z) (1+z)^{-1} \frac{\Theta}{4\pi}\Delta V(z),
\end{equation}

\noindent where $t_c$ is the ``control time'' probability function, which takes into account the survey efficiency with difference magnitude (described in \S 2), $\Theta$ is the area surveyed, and $\Delta V(z)$ is the volume element in a $\Delta z$ shell about $z$ for an assumed flat universe ($\Omega_k=0$)\footnote{The volume is $V(z)=(4\pi/3)D_P^3=(4\pi/3)D_L^3/(1+z)^3$. This is the correct form of the equation, which has a typographical error in S04.}. The expected number of SNe~Ia at $z>1.4$ is then, 

\begin{equation}
        N_{Ia}(z>1.4)=\sum_{z=1.4}^{2.6}N_{Ia}(z)\Delta z. 
\end{equation}

\begin{figure}
        \epsscale{0.8}
        \plotone{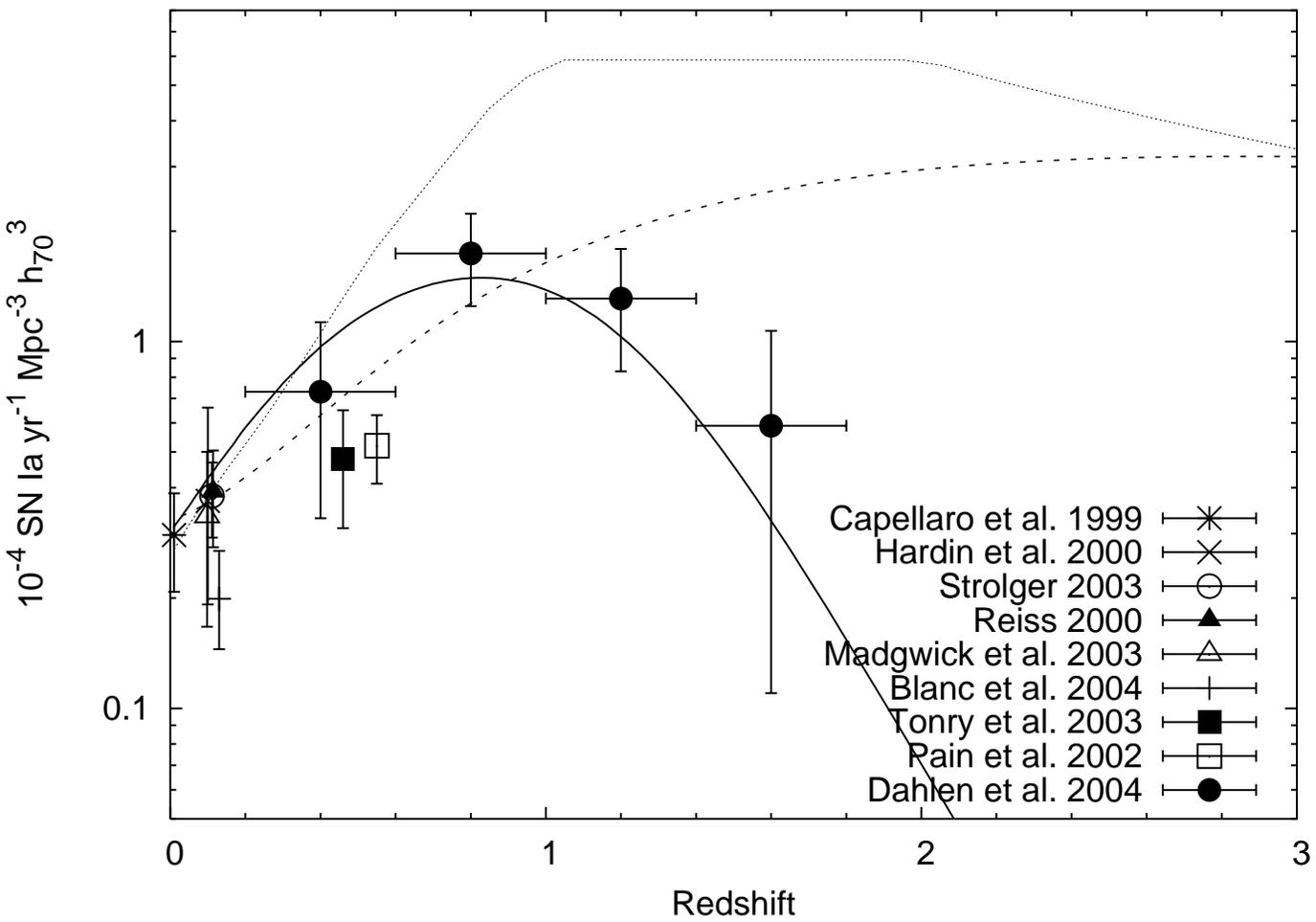}
        \caption{The Type Ia supernova rate history, as measured by several authors (points). Vertical bars represent statistical errors, and horizontal bars on Dahlen et al. (2004) points represent redshift bin size. Also shown are the tested models for the $\mathcal{R}_{Ia}(z)$: The best-fit delayed SFR model (solid line), a SFR-like model (dashed line), and Chary \& Elbaz (2001) SFR model (dotted line). \label{fig:fig4}}
\end{figure}

For comparison, we also calculated the yield expected from a $\mathcal R_{Ia}(z)$ which is directly proportional (i.~e.~without delay) to the extinction corrected SFR$_U$($z$) model used in S04, but with an explosion efficiency for the SN~Ia progenitor mass range of 10\% (instead of 5\%) as required to match SN~Ia rate measurements at $z< 0.2$ by various authors (see Figure~\ref{fig:fig4}). 

\begin{figure}
        \epsscale{0.8}
        \plotone{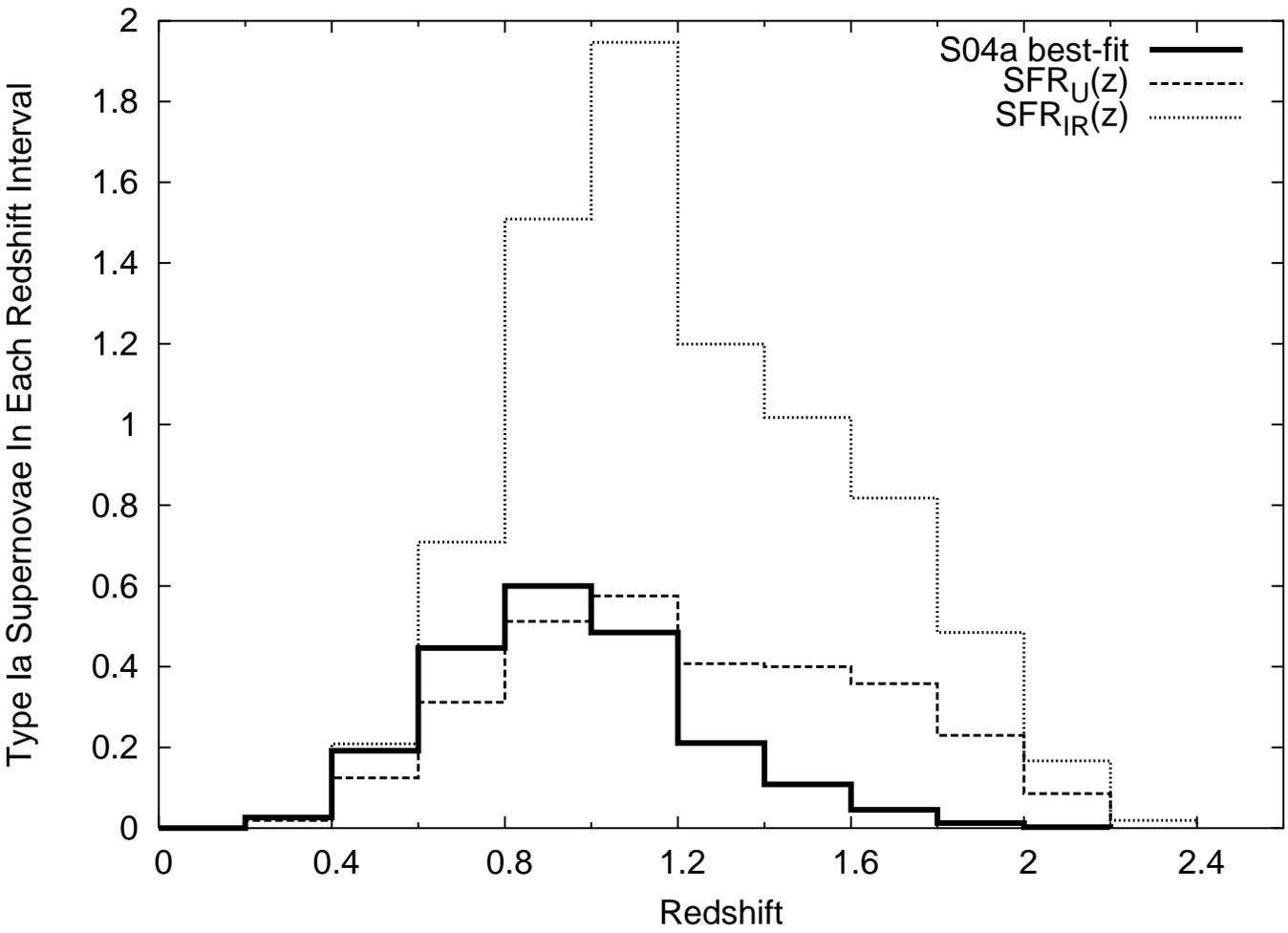}
        \caption{The expected redshift distribution of SNe~Ia for the three tested models in the UDF and UDFP ($F850LP$-band).  \label{fig:fig5}}
\end{figure}

\begin{figure}
        \epsscale{0.8}
        \plotone{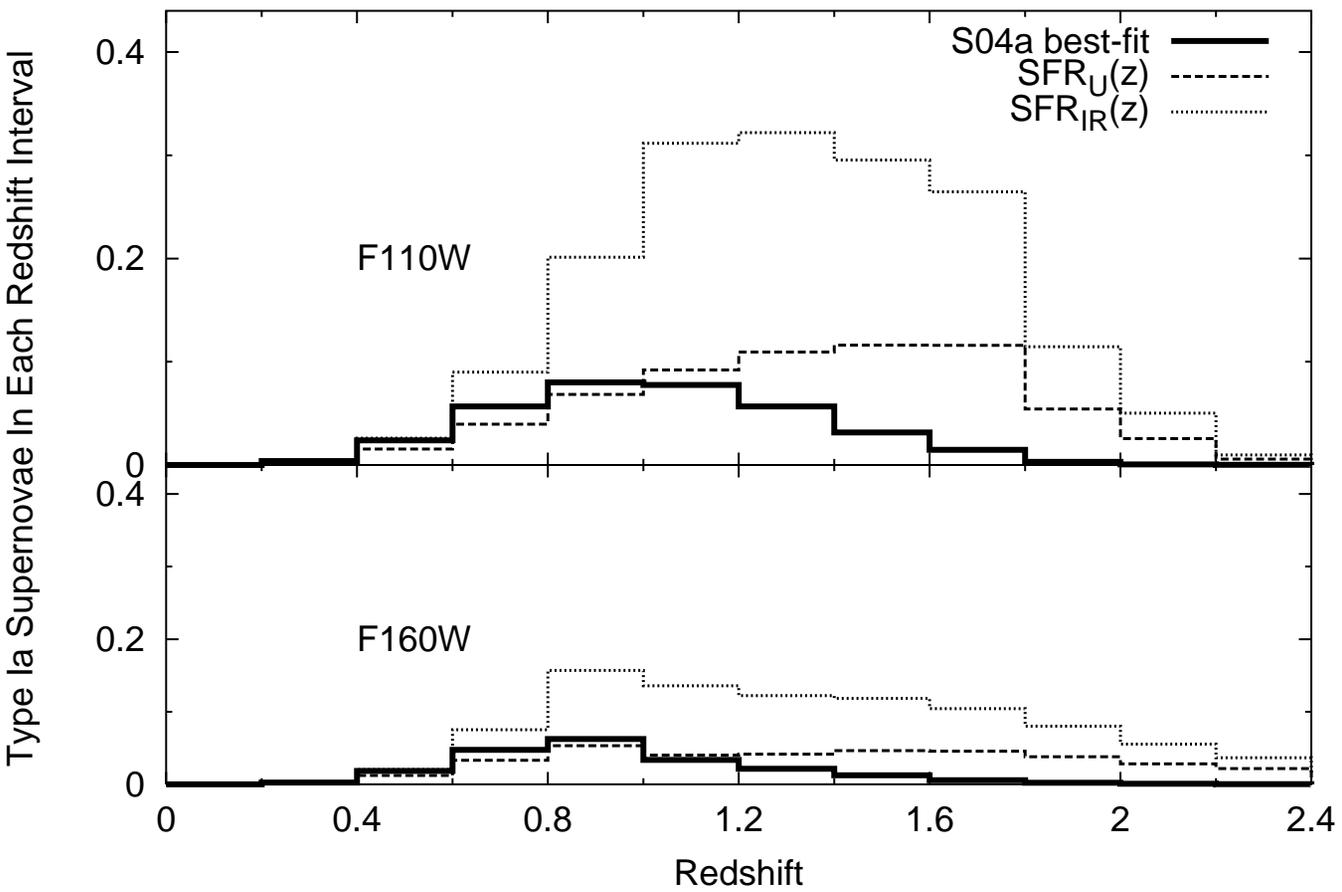}
        \caption{Same as Figure~\ref{fig:fig5} but for the NICMOS UDF in $F110W$ and $F160W$ passbands.\label{fig:fig6}}
\end{figure}

As a straw man, we also computed the yield expected from a $\mathcal R_{Ia}(z)$ which proportional to the SFR$_{IR}$($z$) without any delay where we again assumed the progenitor efficiency to be $10\%$ to match the low-$z$ SN~Ia rates.  The expected number of SNe~Ia from the tested models over all redshift ranges and at $z > 1.4$  for each image stack of the UDF and UDFP (differenced with the prior image stack) are tabulated and summed in Tables~\ref{tab:tab3} and~\ref{tab:tab4}. In Figures~\ref{fig:fig5} and~\ref{fig:fig6} we show the redshift distribution (total number in each redshift bin) expected by each $\mathcal R_{Ia}(z)$ model, in the $F850LP$, $F110W$, and $F160W$-bands.

The best-fit $\mathcal R_{Ia}(z)$ model predicts only 0.06 SNe~Ia at $z>1.4$ in the UDF+UDFP, and 0.02 and 0.01 in the IRUDF $F110W$ and $F160W$-bands respectively,  due to the long delay required by progenitors in the model. The universe was only 4.5 billion years old at $z=1.4$ (assuming $H_o=70$ km s$^{-1}$ Mpc$^{-3}$, $\Omega_m=0.3$, and $\Omega_{\Lambda}=0.7$), and few early forming progenitors would have had sufficient time to explode. We determined the significance of the yield given the model using the Poisson probability for observing a number of events,

\begin{equation}
        P_{\mu}(x)=\frac{\mu^{x}e^{-\mu}}{x!},
\end{equation}
\noindent where $x$ is the integer number of observed SNe~Ia ($x=0$ in this survey) and $\mu$ is the expected number of events determined from the model. The absence of SN~Ia events from this survey in the tested redshift range is an expected result for this best-fit $\mathcal R_{Ia}(z)$ model, with Poisson probabilities of 94.2\%, 98.2\%, and 99.1\% assigned to this outcome in the $F850LP$, $F110W$, and $F160W$-bands, respectively. However, a null result could also be expected from a $\mathcal R_{Ia}(z)$ proprotional to the SFR$_U$($z$) (i.e., with no delay) with Poisson probabilities of 50.4\%, 81.7\%, and 86.7\% for zero observed events in the $F850LP$, $F110W$, and $F160W$-bands.  Although the greatest expected yield would come from the model proportional to the SFR$_{IR}$($z$), a null yield could still be expected in the tested redshift range with probabilities of 22.5\%, 64.4\%, 74.8\% in the respective passbands. 

It is interesting to consider what the relative agreement with the tested models would have been had a single SN~Ia at $z > 1.4$ been discovered in each search. If one high-$z$ SN~Ia had been found in each the $F110W$ and $F160W$-bands, then it would have been very unlikely that S04 best-fit $\mathcal R_{Ia}(z)$ represents the true SN~Ia rate history, as such an outcome would be expected less than 2\% of the time. The S04 best-fit $\mathcal R_{Ia}(z)$ could therefore be rejected to $>98\%$ confidence. However, one SN may have been expected in each IRUDF passbands for either the SFR$_U$($z$) or SFR$_{IR}$($z$) models, although with Poisson expectation probabilities of $12-28\%$, it would appear to be a low likelihood. Similarly, had one SN~Ia at $z>1.4$ been discovered in the $F850LP$ surveys, then S04 best-fit $\mathcal R_{Ia}(z)$ could be rejected at $>94\%$. But again the yield would be only marginally expected from either of the other two models, with a Poisson probability of $\sim 35\%$. 

Although it appears as there was a statistical preference by the data for the best-fit $\mathcal R_{Ia}(z)$ model, it is clear that our null result cannot reject any of the tested models to a significant (greater than 99\%) confidence. In fact, a model would need to predict five or more SNe~Ia at $1.4<z<2.4$ to be rejected to greater than 99\% confidence by the zero yield of this survey. Such a hypothetical model suggests that the average SN~Ia rate in this high-$z$ range would be approximately 150 times larger than the SN~Ia rate measured in the local ($z\le0.1$) universe. This is very inconsistent with trends inferred from measured SN~Ia rates over any redshift interval, and is likely to be rejected based on theoretical considerations of the low [O/Fe] ratios it would predict for stars and the ISM of local galaxies.

Such an increase in the SN~Ia rate is plausible in the SN~Ia rate models considered by~\citet{2004MNRAS.347..951M} in clusters of galaxies. Using $e-$folding delay time functions, described in detail in~\citet{1998MNRAS.297L..17M} and~\citet{1999AA...350..349D}, with $e-$fold times of less than 1 Gyr, one could expect a SN~Ia rate as high as 45 SNu in galaxy clusters at $z\approx1.5$, following a burst of star formation in these clusters at $z=2$. However, to be consistent with the observed iron masses of galaxy clusters (assuming most of this mass is attributed to SNe~Ia), the SN~Ia rate history would likely need to drop to zero at $z < 0.7- 1.0$. This would be inconsistent with the non-negligible SN~Ia rate measurements in galaxy clusters at $0.04 < z < 0.08$~\citep{reiss2000}. 

There is significant evidence for uncertainty in the estimates of galaxy volume densities for specific populations in the GOODS fields~\citep{2004ApJ...600L.171S}, and the survey presented in this paper covers a smaller area of the GOODS South field. However, the SFR$_U$($z$) model (in the $z>1$ regime) used in this analysis was also determined from GOODS data, and therefore should be subject to the same variances. The SFR$_{IR}$($z$) model, by contrast, was determined from several local surveys, and through theoretical modeling of the evolution of the galaxy infrared luminosity function~\citep{2001ApJ...556..562C}. Therefore, we expect cosmic variance to distort the predicted SN~Ia rate in this observed volume (from $15 - 20\%$) when applying the global average SFR$_{IR}$($z$) to this small field.

It should be noted that our lower-bound on the redshift range tested, $z>1.4$, was based on the redshift range where the completeness of the UDF+UDFP survey becomes significant over the GOODS SN survey.  However, our results are not very sensitive to the precise definition of the uniquely high-redshift space of this survey. Additionally, other baselines and combinations of images in the UDF and UDFP surveys could have been selected which would have increased the expected number of SNe Ia in the desired redshift range. However, an increase by a factor of 5 or more would be necessary to begin to significantly rule out some of the tested models. Tests of varying cadences between visits and combining image stacks result in changes in the expected yield of the tested models of less than a factor of two in the considered redshift range.

It appears that deepest-field-type, pencil beam surveys (i.e., the UDF or the Hubble Deep Field) with {\it HST} lack the volume necessary to accumulate a statistically significant sample to differentiate between the current plausible $\mathcal R_{Ia}(z)$ models in the redshift regimes in which they are uniquely sensitive. This is not surprising, considering that the increased exposure times over the GOODS did not significantly increase the luminosity distance probed. It would be necessary to repeat the UDF survey several times (or an increase of several times the survey area) in order to achieve expected yields for significant Poisson probability rejection of the tested models. This would not be economical with {\it HST} given the enormity of the UDF survey.

Progress in determining the rates of SNe Ia and their associated
timescales for assembly is most likely to come by ``piggybacking''
such studies on the now numerous and major
efforts to find and measure supernovae to constrain
dark energy parameters.  Such ``cosmology-driven'' SN surveys are
designed as wide, open field surveys and are well-suited to deriving
the supernova field rates.  These surveys generally 
avoid galaxy clusters as search targets
because half of the yield of SNe Ia found in such surveys are located 
behind the clusters~\citep{2002MNRAS.332...37G, 1998AJ....115...26R} where contamination 
by lensing is unavoidable.  In addition, potential evolutionary differences
between cluster and field-born SNe Ia could add 
systematic errors to cosmological determinations.  As a result, very deep rates 
and cluster rates of SNe Ia have been
studied on a much smaller scale than field rates.

\section{Summary}

The UDF, UDFP, and IRUDF observations yielded four SNe over the 5 month duration of the survey. We find the dearth of SNe Ia at high redshift in this survey to be consistent with the best-fit $\mathcal R_{Ia}(z)$ from S04. However, it is also a likely result from a $\mathcal R_{Ia}(z)$ which follows the SFR$_U$($z$) at a higher progenitor explosion efficiency. 
More rapidly increasing $\mathcal R_{Ia}(z)$, such as the~\citet{2001ApJ...556..562C} SFR$_{IR}$($z$) model without delay, also cannot be significantly rejected by these data, because of the relatively small area probed in this survey. The general conclusion is that these data are well supported by the GOODS best-fit model, which shows that SNe~Ia progenitors require substantial (approximately 4 Gyr) delays from star formation to produce events. However, other plausible SN~Ia rate models cannot be excluded. Surveys such as the GOODS more than compensate for the loss in depth by the increased survey area, and are vastly superior programs for differentiating between SN~Ia rate models. Future endeavors with {\it HST} such as the Cycle 12 programs by A. Riess and S. Perlmutter (PIs for GOs 9727 and 9728 respectively) and the {\it Probing Acceleration Now with Supernovae} (or PANS, GO-10189, A. Riess, PI) in Cycle 13 will allow for much more precise measure of the SN~Ia rate in the $z>1$ regime. In combination with the GOODS data, they will provide the best available measures of the empirical distributions of SN~Ia delay times and the comparison to star formation rate models.

We thank Tomas Dahlen for his beneficial discussions. We also thank Steven Beckwith, Roger Thompson, Bahram Mobasher, Anton Koekemoer, Louis Bergeron, Rogier Windhorst, and Rychard Bouwens. Financial support for this work was provided by NASA through programs GO-9728, GO-9978, and GO-9803 from the Space Telescope Science Institute, which is operated by AURA, Inc., under NASA contract NAS 5-26555.

\begin{deluxetable}{cccccccccc}
        \tabletypesize{\scriptsize} 
        \tablewidth{0pc}
        \tablecaption{Predicted yield of SNe~Ia from UDF+UDFP\label{tab:tab3}}
        \tablehead{
        \colhead{Field}&\colhead{Mean Date}&\colhead{\# of}&\colhead{Exp. Time}&\colhead{5$\sigma$ lim.}&\colhead{Baseline}&\multicolumn{3}{c}{$\mathcal R_{Ia}(z)$ models}\\
        \colhead{}&\colhead{}&\colhead{Orbits}&\colhead{(seconds)}&\colhead{(mag)}&\colhead{(days)}&\colhead{S04 best-fit}&\colhead{SFR$_U$}&\colhead{SFR$_{IR}$}}
        \startdata

GOODS&2003 Feb 11&5&10500&\nodata&\nodata&\nodata&\nodata&\nodata\\
0924-1002&2003 Sep 28&18&43740&27.3&229&1.013 (0.017)&1.256 (0.206)&3.425 (0.444)\\
1002-1008&2003 Oct 05&18&43740&27.4&7&0.031 (0.002)&0.065 (0.023)&0.169 (0.050)\\
1010-1014&2003 Oct 12&18&43740&27.4&7&0.033 (0.003)&0.076 (0.032)&0.195 (0.071)\\
1016-1029&2003 Oct 22&20&48600&27.5&11&0.052 (0.005)&0.129 (0.059)&0.326 (0.128)\\
1204-1211&2003 Dec 07&16&38480&27.3&46&0.190 (0.014)&0.429 (0.173)&1.097 (0.373)\\
1212-1218&2003 Dec 15&18&41940&27.4&8&0.035 (0.002)&0.075 (0.027)&0.195 (0.058)\\
1224-1230&2003 Dec 27&12&29160&27.1&12&0.049 (0.004)&0.105 (0.038)&0.272 (0.084)\\
0101-0111&2004 Jan 06&8&19440&26.9&10&0.040 (0.002)&0.076 (0.022)&0.202 (0.049)\\
0113-0114&2004 Jan 13&6&14580&26.7&8&0.032 (0.002)&0.057 (0.015)&0.153 (0.033)\\
&&&&&{\bf UDF =}&{\bf1.475 (0.051)} & \bf{2.268 (0.595)} & \bf{6.034 (1.290)}\\
\tableline
GOODS&2003 Feb 11&2&4200&\nodata&\nodata&\nodata&\nodata&\nodata\\
0831-0901&2003 Aug 31&2&4600&26.1&202&0.595 (0.006)&0.651 (0.056)&1.766 (0.125)\\
0911-0913&2003 Sep 12&9&20700&26.8&12&0.058 (0.003)&0.114 (0.034)&0.302 (0.075)\\
&&&&&{\bf UDFP = }&{\bf 0.653 (0.009)} & \bf{0.765 (0.090)} & \bf{2.068 (0.200)}\\
\tableline
\tableline
&&&&&{\bf TOTAL = }&{\bf 2.128 (0.060)}&\bf{3.033 (0.685)}&\bf{8.100 (1.490)}&

\tablecomments{The expected number of SNe~Ia in the UDF+UDFP for each model per template-search pair are shown over all redshifts, and with the expected number at $z>1.4$ shown in parentheses. The numbers are summed for each survey field (in bold), and for the entire survey.}
        \enddata
\end{deluxetable}

\begin{deluxetable}{cccccccccc}
        \tabletypesize{\scriptsize} 
        \tablewidth{0pc}
        \tablecaption{Predicted yield of SNe~Ia from NICMOS UDF\label{tab:tab4}}
        \tablehead{
        \colhead{Field}&\colhead{Mean Date}&\colhead{\# of}&\colhead{Exp. Time}&\colhead{5$\sigma$ lim.}&\colhead{Baseline}&\multicolumn{3}{c}{$\mathcal R_{Ia}(z)$ models}\\
        \colhead{}&\colhead{}&\colhead{Orbits}&\colhead{(seconds)}&\colhead{(mag)}&\colhead{(days)}&\colhead{S04 best-fit}&\colhead{SFR$_U$}&\colhead{SFR$_{IR}$}}
        \startdata
$F110W$\\
Stack 1&2003 Sep 07&4&10752&\nodata&\nodata&\nodata&\nodata&\nodata\\
Stack 2&2003 Nov 23&4&10752&25.6&77&0.341 (0.017)&0.605 (0.170)&1.598 (0.374)\\
\tableline
$F160W$\\
Stack 1&2003 Sep 07&4&10752&\nodata&\nodata&\nodata&\nodata&\nodata\\
Stack 2&2003 Nov 23&4&10752&24.2&77&0.234 (0.012)& 0.481 (0.216)&1.166 (0.429)\\

\tablecomments{Same as in Table~\ref{tab:tab3}, but for the IRUDF.}
        \enddata
\end{deluxetable}

\begin{deluxetable}{ccccccc}
        \tablewidth{0pc}
        \tablecaption{Positional Data\label{tab:tab1}}
        \tablehead{\colhead{SN}& \colhead{R. A. (2000)}& \colhead{Decl. (2000)}& \colhead{N (arcsec)}& \colhead{E (arcsec)}& \colhead{Redshift}}
        \startdata
K0302-001&03:32:37.10&-27:56:53.6&-0.27&0.13&1.3 ($1.2-1.4$)\\
2003lt&03:32:42.88&-27:55:52.5&0.45&0.20&1.0 ($0.74-1.26$)\\
2003lu&03:32:36.17&-27:55:01.4&-0.28&0.43&0.11 ($0.0 - 0.24$)\\
2004R&03:32:41.30&-27:46:13.6&-0.10&-0.11&0.80 ($0.56-1.04$)\\
        \enddata
        \tablecomments{Offsets are from the nucleus of the host galaxy to the SN. Maximum likelihood photometric redshifts are listed with the 95\% confidence intervals in parentheses. Coordinates supersede those announced in the IAU Circular.} 
\end{deluxetable}

\begin{deluxetable}{cccc}
        \tablewidth{0pc}
        \tablecaption{Photometric Data\label{tab:tab2}}
        \tablehead{\colhead{SN}& \colhead{Filter}& \colhead{JD+ 2,450,000}& \colhead{Magnitude}}
        \startdata
K0302-001&$F850LP$&2680.22&24.38 (0.03)\\
&$F775W$&2680.96&25.20 (0.03)\\
&$F606W$&2680.25&27.71 (0.12)\\
\tableline
2003lt&$F850LP$&2890.58&25.85 (0.06)\\
&&2895.71&26.19 (0.04)\\
&$F775W$&2885.32&27.26 (0.11)\\
&$F606W$&2887.52& 30.2 (0.8)\\
\tableline
2003lu&$F850LP$&2895.71&26.33 (0.04)\\
\tableline
2004R&$F850LP$&3006.95&27.54 (0.26)\\
&&3014.59&26.79 (0.13)\\
&&3015.49&26.87 (0.10)\\
&&3017.13&26.49 (0.06)\\
&$F775W$&3005.89&27.91 (0.26)\\
&&3009.21&27.09 (0.12)\\
&&3013.70&27.48 (0.07)\\
&&3017.45&27.12 (0.05)\\
&$F606W$&3000.40&27.82 (0.07)\\
&&3007.75&27.85 (0.06)\\
&&3016.28&28.04 (0.08)\\
&&3019.15&28.20 (0.09)
        \enddata
        \tablecomments{Magnitudes are given in the Vega-based system and are listed with their photometric errors in parentheses.}
\end{deluxetable}

\end{document}